\documentclass[pre,floatfix,superscriptaddress,twocolumn,showpacs]{revtex4-1}
\usepackage{graphicx}
\usepackage{amsmath,amssymb,amsfonts}
\usepackage{mathtools}
\usepackage{bm}
\usepackage{color}
\usepackage[colorinlistoftodos]{todonotes}

\begin{document}
\title{Density and inertia effects on two-dimensional active semiflexible filament suspensions}
\author{Giulia Janzen}
\author{D.A. Matoz-Fernandez}
\email{dmatoz@ucm.es}
\affiliation{Department of Theoretical Physics, Complutense University of Madrid, 28040 Madrid, Spain}
\date\today
\begin{abstract}
We examine the influence of density on the transition between chain and spiral structures in planar assemblies of active semiflexible filaments, utilizing detailed numerical simulations. We focus on how increased density, and higher P\'eclet numbers, affect the activity-induced transition spiral state in a semiflexible, self-avoiding active chain. Our findings show that increasing the density causes the spiral state to break up, reverting to a motile chain-like shape. This results in a density-dependent reentrant phase transition from spirals back to open chains. We attribute this phenomenon to an inertial effect observed at the single polymer level, where increased persistence length due to inertia has been shown in recent three-dimensional studies to cause polymers to open up. Our two-dimensional simulations further reveal that a reduction in the damping coefficient leads to partial unwinding of the spirals, forming longer arms. In suspension, interactions among these extended arms can trigger a complete unwinding of the spirals, driven by the combined effects of density and inertia.
\end{abstract}
\maketitle
\section{Introduction}
Spiral structures in nature, such as those observed in the shells of ancient cephalopods, represent a ubiquitous pattern across numerous biological systems~\cite{Eldredge1996, Barabe1998, liu2024}. The occurrence of spiral symbolism in archaeological finds from various civilizations is proof that these patterns not only highlight biological diversity but also resonate with the growth of human culture~\cite{o1984newgrange}. Golden spirals and related proportions are frequently used in architectural designs to establish harmonious relationships between structural and functional elements. Examples of practical applications of these ratios are found in Le Corbusier's \textit{Modulor} and van der Laan's \textit{Plastic Number}~\cite{bangs2006return, meisner2018golden}. Historical attempts to decode the mechanics of spiral structures date back to antiquity, with Archimedes famously developing a foundational method for their geometric construction~\cite{Archimedes225BC}.

The formation of these spirals can be attributed to processes occurring in systems far from thermodynamic equilibrium, providing a rich area for theoretical modeling \cite{kuramoto1984chemical, cross1993pattern, mikhailov2006control}. Examining the widespread occurrence of these structures has revealed that comparable geometric patterns also extend into the linear filamentous structure configurations found in both natural and artificial systems ranging from the microscopic world such as microtubules and actin filaments in the cytoskeleton of the cell~\cite{bray2000cell, Schaller2010, ganguly2012cytoplasmic}, to bacteria~\cite{Yaman2019}, to worms~\cite{deblais2020phase, deblais2020rheology, nguyen2021emergent}, to granular systems~\cite{wen2012polymerlike, soh2019self}, to macroscopic robotic systems~\cite{marvi2014sidewinding, ozkan2021collective}, not only illustrate a continuity of different pattern motifs but also underscore the universal application of underlying physical principles. This convergence suggests that filamentous extensions, spiral geometries, and other complex structures share common mechanistic foundations, potentially governed by the same far-from-equilibrium processes~\cite{Ravichandran2017, Vliegenthart2020, li2023nonequilibrium, Locatelli2024}. To study such systems, in recent years, it has become increasingly clear that active matter provides a powerful framework for understanding the physics of living systems across scales. This includes diverse scales ranging from components inside the cell~\cite{Schaller2010, Sumino2012}, to artificial self-propelled particles ~\cite{Narayan2007, Theurkauff2012, Palacci2013}, to microorganisms ~\cite{Saw2017, Beer2019},  to large animal species ~\cite{Sumpter2010, Bialek2012}, to robot swarms ~\cite{Rubenstein2014, Wang2021}, and what filamentous matter~\cite{IseleHolder2015, Winkler2020, philipps2022tangentially, ubertini2024universal, malgaretti2024coil}. 

\begin{figure}[t!]
    \centering
    \includegraphics[width=8.5cm]{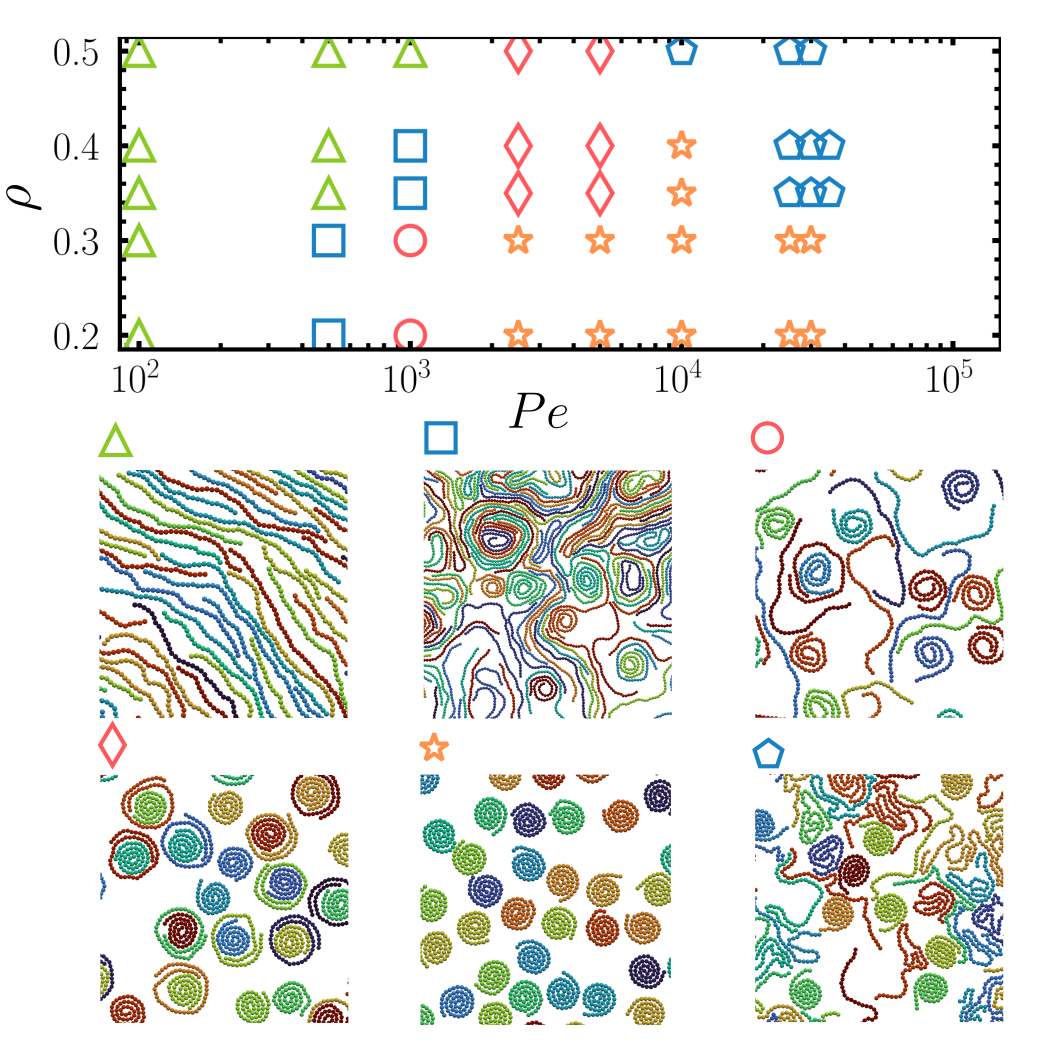}
     \caption{Phase diagram in the $(\rho,Pe)$-plane, computed from the probability distribution of the turning number $P(\psi)$. Different phases are represented by distinct symbols: stable open chains without spirals (green triangles),  stable chains with metastable loops and spirals (blue squares), stable open chains coexisting with stable spirals (red dots), spirals coexisting with metastable loops (red diamonds),  stable spirals (orange stars), and stable compact chains with metastable spirals (blue pentagons). At low density, the system exhibits an open chains phase for low $Pe$, transitions to a phase with coexisting spirals and open chains around $Pe \approx 10^3$, and ultimately reaches a spiral state at $Pe \approx 2.5 \times 10^3$. At intermediate density ($0.3 < \rho < 0.4$), a similar transition as the one observed at low density, occurs from open chains to spirals, followed by a reentrant phase for $Pe > 10^4$, characterized by the coexistence of compact open chains and spirals. At high densities ($\rho > 0.4$), the transition from open chains to compact chains and spirals happens at $Pe \approx 10^4$. However, due density induced steric effects, the system never reaches a state in which all filaments are spirals.}
\label{fig:1}
\end{figure}

In this paper, we investigate the effect of density on the chain-to-spiral transition in planar assemblies of active semiflexible filaments, through detailed numerical simulations~\cite{IseleHolder2015, Prathyusha2018}. We specifically explore how density influences the activity-induced transition into a spinning spiral state within a semiflexible, self-avoiding chain. Our results reveal that increasing the density, marked by increased P\'eclet numbers, the spiral state becomes destabilized, resulting in a motile chain-like structure. This leads to a density-dependent reentrant phase transition from spirals back to open chains (Fig.~\ref{fig:1}). This phenomenon can be understood as an inertial effect at the single polymer level, where recent three-dimensional studies have shown an increase in persistence length due to inertia, causing polymers to open up~\cite{Fazelzadeh2023}. Consistent with these findings, we observe in two dimensions that a reduction in the damping coefficient causes the spirals to partially unwind, forming longer arms. When these chains are placed in suspensions, the interaction between these extended arms can lead to a complete unwinding of the spirals, driven by the combined effects of density and inertia (Fig.~\ref{fig:6}).

\section{Model} 
We follow closely the model used in Ref.~\citeauthor{Prathyusha2018}~\cite{Prathyusha2018}. The system is composed of $N_f$ filaments, each of which with $N_b$ beads. In this paper, we ignore long-range hydrodynamic interactions, and thus, we consider the dry limit, where the damping from the medium dominates so that the surrounding fluid only provides single-particle friction. In this regime, the equations of motion for each filament are given by the Langevin equation~\cite{IseleHolder2015, Prathyusha2018},
\begin{equation}
m_i \, \ddot{\mathbf{r}_i}=-\gamma \, \dot{\mathbf{r}_i}+\mathbf{R}_i(t)+\sum_{j\neq i } \mathbf{f}_{ij}+\mathbf{f}^a_i,
\end{equation}
where $m_i$ is the mass of bead $i$, $\mathbf{r}_i=(x_i,y_i)$ represents the bead's spatial coordinates and the dot denotes the time derivatives, $\gamma$ is the damping coefficient and $\mathbf{R}_i(t)$ is a delta-correlated random force with zero mean and variance   
$4 \gamma k_B T \delta_{ij} \delta(t-t^\prime)$, where  $k_B$ is the Boltzmann constant and $T$ the temperature. Moreover, $\mathbf{f}_{ij}=-\nabla_i \phi(r_{ij})$ is the interaction force between beads $i$ and $j$, where $r_{ij}=\left\vert\textbf{r}_i-\textbf{r}_j \right\vert$ and $\phi$ is the bonded and short-range nonbonded pair potentials $\phi=\phi_B+\phi_{NB}$. Bonded interactions $\phi_B=\phi_{bond}+\phi_{bend}$ account for both chain stretching, modeled by the Tether bond potential~\cite{Noguchi2005} and bending modeled with the harmonic angle potential~\cite{Prathyusha2018}.
Note that in the limit of small bond-length variations, which is the one that we consider here, the use of Tether or FENE bond potential should not influence the results. Next, the nonbonded interactions, $\phi_{NB}$ account for steric repulsion and are modeled with the Weeks-Chandler-Anderson (WCA) potential~\cite{weeks1971role}. All of the details for the bonded and nonbonded interaction parameters are described in the Supplementary Material~\cite{SI}.

Lastly, $\mathbf{f}^a_i$ is the active self-propulsion force on bead $i$~\cite{jiang2014motion}
\begin{equation}
   \mathbf{f}^a_i= f_a (\mathbf{t}_{i-1,i}+\mathbf{t}_{i,i+1}), 
\end{equation}
where $f_a$ is the magnitude of the active force, and $\mathbf{t}_{i,i+1}=\mathbf{r}_{i,i+1}/r_{i,i+1}$ is the unit-length tangent vector along the bond which connects beads $i$ and $i+1$, the end beads have contributions from only one neighboring bond. It should be noted that various definitions exist for the active force on particle $i$ as discussed in recent literature (see~\cite{Bianco2108, jiang2014motion, IseleHolder2015}), all of which converge to the same continuum limit~\cite{Winkler2020}. In simulations of models~\cite{jiang2014motion, IseleHolder2015}, deviations may occur when the activity surpasses a certain threshold, especially in cases with weak bond potentials and/or with significant bond-length variations, or when the number of monomers is small.

In our study, we set the degree of polymerization at \(N_b = 50\), monomer size \(\sigma = 1.0\), interaction energy scale \(\epsilon = 1.0\), thermal energy \(k_B T = 10^{-1}\), and damping coefficient $\gamma = 1.0$. {The length of each active polymer was calculated as \({L}\approx0.96(N_b-1)\sigma\)}. We defined two critical dimensionless numbers for our analysis: the scaled persistence length \(\xi_p/L = 2b\kappa/(Lk_B T)\), which reflects the filament's thermal (passive) stiffness, and the active P\'eclet number, \(Pe = f_p L^2 / (\sigma k_B T)\), which characterizes the dominance of active forces over thermal fluctuations. Simulations were conducted using the GPU-enhanced SAMOS molecular dynamics package~\cite{SAMoS2024}. The BAOAB Langevin scheme was employed for numerical integration~\cite{leimkuhler2015molecular}. Analysis was performed using custom Python scripts, and visualizations were generated using the ParaView~\cite{fabian2011paraview}. To explore the phase behavior of the system we set the number of filaments $N_f = 10^{3}$, vary the packing fraction $\rho = N_f N_b \,\pi\,\sigma^2/4L_{box}^{2}$, and the P\'eclet number ($Pe$). For further details, refer to the Supplementary Material~\cite{SI}.
\section{Results and discussion}
\subsection{Collective Static Behavior}
\label{sec:collective}
\begin{figure}[t!]
    \centering
    \includegraphics[width=8.5cm]{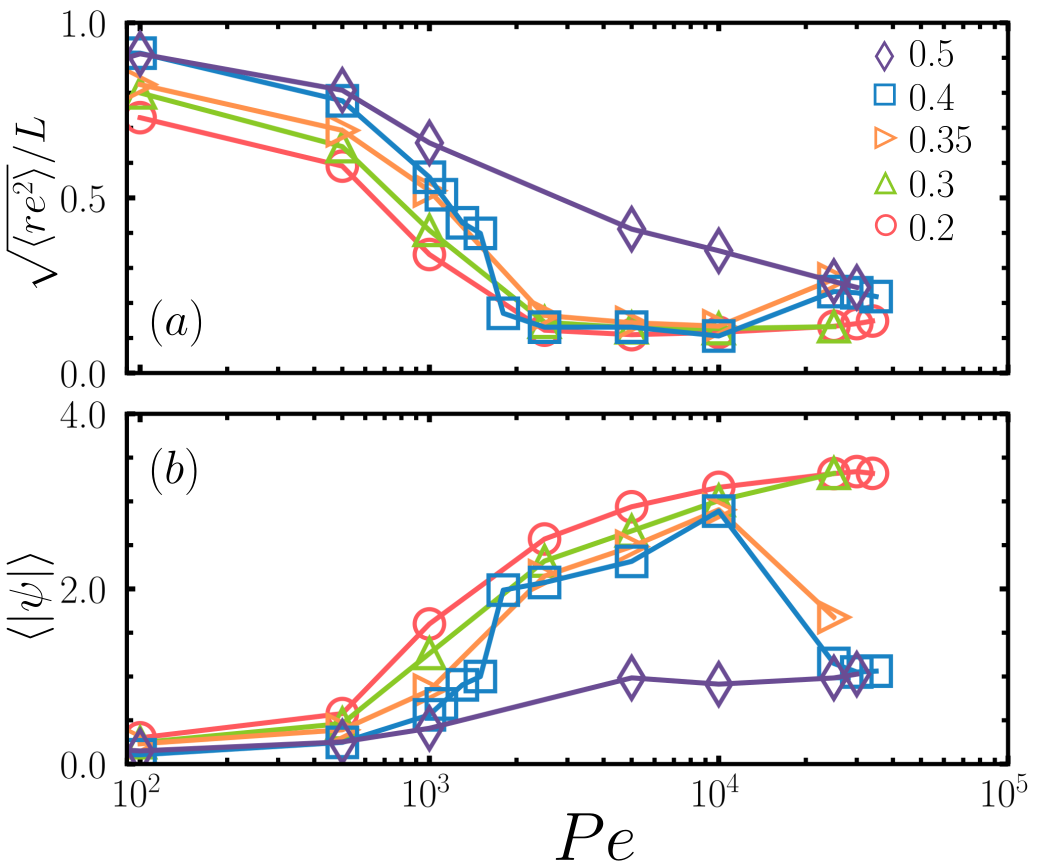}
\caption{Structural properties of the filaments, averaged over the number of filaments $N_f$, as a function of the P\'eclet number $Pe$ for different densities represented by distinct symbols and colors: purple diamonds for $\rho=0.50$, blue squares for $\rho=0.40$, orange right triangles for $\rho=0.35$, green triangles for $\rho=0.30$, and red dots for $\rho=0.2$.
(a) The end-to-end distance, $\sqrt{\langle r^2_e \rangle}/L$. At low densities ($\rho \leq 0.30$), the end-to-end distance decreases until approximately $Pe \approx 2.5 \times 10^4$, where it stabilizes around $\approx 10^{-1}$, corresponding to the typical end-to-end distance of spirals. For intermediate densities ($0.3 < \rho \leq 0.4$), the end-to-end distance behaves similarly to that observed at low density. However, for $Pe>10^4$, a reentrant phase occurs due to spirals opening up, increasing the end-to-end distance. Finally, at high density ($\rho=0.5$), the absence of a phase where all filaments are spirals leads to a monotonic decrease in the end-to-end distance as a function of $Pe$.
(b) The turning number $\langle \vert\psi\vert \rangle$, defined by Eq.(\ref{eq:turning_number}). At low density, the turning number $\langle \vert\psi\vert \rangle$ increases until it plateaus around $Pe \approx 2.5 \times 10^3$, where $\langle \vert\psi\vert \rangle \approx 3$, typical of spiral structures. Intermediate densities exhibit a similar trend in $\langle \vert\psi\vert \rangle$ as observed at low densities up to $Pe \approx 10^4$, after which the turning number decreases due to the reentrant phase. At high density, the turning number remains almost constant for all $Pe$ values due to the absence of a pure spiral phase.}
    \label{fig:2}
\end{figure}
To characterize the behavior of the system across different densities and P\'eclet numbers, we analyze two averaged structural properties: end-to-end distance and the turning number.
The first structural property that we compute is the end-to-end distance, denoted by $\sqrt{\langle r^2_e \rangle}/L$, with $r_e = \lVert\mathbf{r}(N_f) - \mathbf{r}(0) \rVert$, where $\mathbf{r}(N_f)$ and $\mathbf{r}(0)$ represent the positions of the last and first monomer of the polymer, respectively.
For a straight filament, $\sqrt{\langle r^2_e \rangle}/L$ is approximately one, as the end-to-end distance coincides with the total length of the polymer. For a circular filament, instead, $\sqrt{\langle r^2_e \rangle}/L$ is approximately zero, and for every other shape falls between 0 and 1.

We found that at low densities ($\rho \leq 0.30$) and small P\'eclet numbers ($Pe<10^2$), the end-to-end distance is approximately one, indicating that all filaments in the system are straight chains (Fig.~\ref{fig:2}(a)). As the P\'eclet number increases, $\sqrt{\langle r^2_e \rangle}/L$ decreases and stabilizes at $\approx 10^{-1}$, showing that the filaments are highly compact. This behavior is consistent with the single-polymer model \cite{IseleHolder2015}. 
Moreover, at intermediate densities ($0.30< \rho \leq 0.40$), the end-to-end distance behaves similarly to that at low density until $Pe=10^4$, when $\sqrt{\langle r^2_e \rangle}/L$ increases again. This phenomenon occurs due to the high P\'eclet number and to excluded volume effects, causing the compact filaments to open up and form chains with larger end-to-end distances than that observed at low density for the same $Pe$. This increase is the signature of the occurrence of the reentrant phase~\cite{Shee2021}.
Lastly, at higher density ($\rho\geq 0.5$), the end-to-end distance starts at one for small P\'eclet numbers, indicating that all filaments are initially open chains. It then decreases monotonically as $Pe$ increases. Despite decreasing, the end-to-end distance never reaches the value of $10^{-1}$ due to the absence of a phase where all filaments are highly compact. The absence of this phase can be attributed to the high density. In this regime, ~\citeauthor{Prathyusha2018}~\cite{Prathyusha2018} identified four distinct phases: melt, flowing melt, swirls, and spirals. 
 
While this end-to-end distance gives information about structural changes, it does not distinguish between shapes. Several filament patterns may cause the first and last monomers to be close together, resulting in comparable end-to-end distances across shapes. To properly describe the various phases, we computed the turning number~\cite{krantz1999handbook,Shee2021} $\psi_i$ for each polymer, where $i$ ranges from 1 to the total number of spirals $N_f$
\begin{equation}
    \psi = \frac{1}{2 \pi} \sum_{j=1}^{N_b-1} (\theta_{j+1}-\theta_j),
\label{eq:turning_number}
\end{equation}
where $(\theta_{j+1}-\theta_{j})$ represents the angle's variation between two consecutive monomers. Therefore, the turning number $\psi$ quantifies the number of turns the chain makes between its two ends. For a straight chain $\psi=0$; for a circle $\psi=1$; and larger values of $\psi$ correspond to loops or spirals. 
At low densities, and intermediate P\'{e}clet number ($Pe=10^3$), the turning number exceeds 1, indicating the occurrence of spirals and loops, in addition to open chains (Fig.~\ref{fig:2}(b)). As $Pe$ increases, the turning number rises until it reaches a plateau for $|\psi| \approx 3.5$, corresponding to that observed for spirals. This indicates that all the filaments in the system are spirals.  
Moreover, at intermediate densities,  the turning number follows a similar trend as observed at low densities up to $Pe \approx 10^4$. Beyond this value, the turning number decreases due to the emergence of the reentrant phase. 
 This behavior in the turning number mirrors that of the end-to-end distance. However, the presence of the reentrant phase is more noticeable in the turning number, as the difference between compact chains and spirals is more significant compared to the difference in end-to-end distance.
Finally, at high density, the turning number remains nearly constant for all $Pe$ values due to the absence of a pure spiral phase.

In summary, the structural properties, show that the system undergoes distinct phases depending on its density. At low density, a single-phase transition occurs from open chains to spirals. In contrast, at intermediate densities, following the transition from open chains to spirals, the averaged properties exhibit a non-monotonic behavior, indicating the emergence of a reentrant phase. Finally, at higher densities, the system is in the open chain phase at small $Pe$, while at higher P\'eclet numbers, spirals begin to emerge. 

\subsection{Characterizing the Polymer's Shape}
\begin{figure}[t!]
    \centering
    \includegraphics[width=8.5cm]{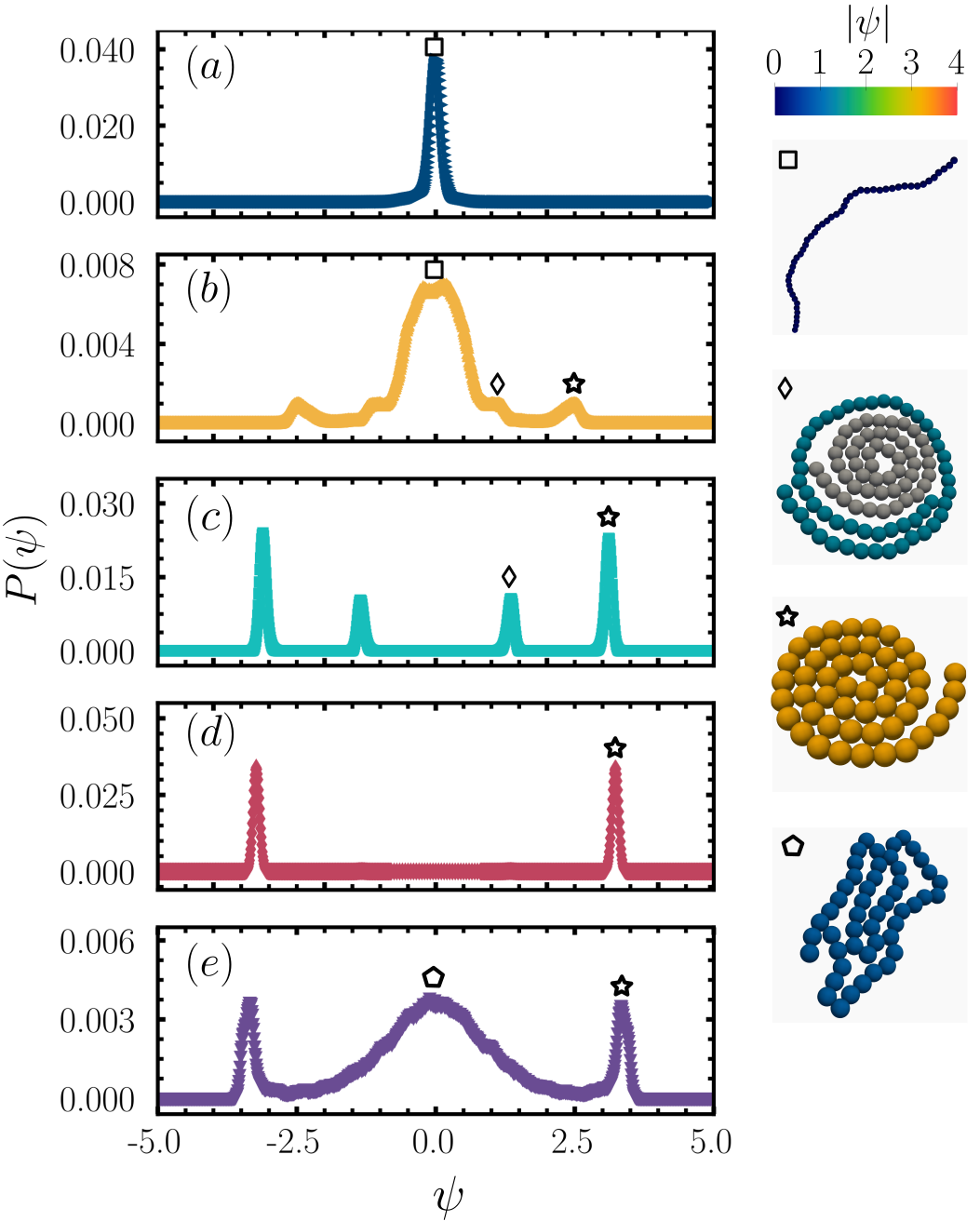}
     \caption{Probability distribution of the turning number $P(\psi)$ at density $\rho=0.4$. (a) The right blue triangles represent $P(\psi)$ for $Pe=10^2$, with black squares indicating the shape of the filaments at the peak where $\psi=0$. As illustrated in the inset, all filaments are stable open chains. (b) The yellow triangles correspond to $Pe=10^3$. In this scenario, $P(\psi)$ reveals five peaks: the highest peak (black squares) represents stable open chains; two peaks for $1<\vert\psi\vert<2$ signify metastable loops (black diamonds); and two peaks for $2<\vert\psi\vert<3$ indicate metastable spirals (black stars). (c) Cyan squares represent $Pe=5.0 \times 10^3$, showcasing four peaks in $P(\psi)$: two for $1<\vert\psi\vert<2$ corresponding to metastable loops (black diamonds), and two for $2<\vert\psi\vert<3$ indicating stable spirals (black stars). (d) Red diamonds correspond to $Pe=10^4$, displaying two peaks for $3<\vert\psi\vert<4$, denoting stable spirals (black stars). (e) Upside-down purple triangles correspond to $Pe=2.5 \times 10^4$, revealing three peaks in $P(\psi)$: one for $\psi=0$ representing stable compact chains (black pentagons), and two peaks for $3<\vert\psi\vert<4$ representing metastable spirals (black stars). The insets show the filament's shapes corresponding to the peaks in the probability distribution $P(\psi)$. The color code is based on the turning number $\vert\psi\vert$ for each specific shape.}
     \label{fig:3}
\end{figure}
 In the previous section, we showed that at intermediate densities, the system's behavior becomes significantly more complex than in the dilute and high-density cases, therefore, in this section, we will mainly focus on the findings obtained for $\rho=0.4$. 
To properly characterize all the different phases observed at this density, we will focus on the probability distributions of two structural properties that exhibit considerable variation across the different shapes present in the system: the turning number and the curvature. Figure ~\ref{fig:3}(a) shows that the probability of the turning number $P(\psi)$ exhibits a single peak at $\psi=0$ since all filaments in the system are stable open chains. The typical shape of the polymers at this $Pe$ is illustrated in the first inset of Fig.~\ref{fig:3}, marked by a black square.

Increasing the P\'eclet number, we found that $P(\psi)$ presents five distinct peaks: one at $\psi=0$, indicating that the system is predominantly composed of stable open chains; two symmetric peaks for $1<\vert\psi\vert<2$, corresponding to the presence of metastable loops; and finally, two symmetric peaks for $2<\vert\psi\vert<3$, indicating the presence of metastable spirals (Fig.~\ref{fig:3}(b)). Thus, the system exhibits a coexistence of three different shapes. 
Increasing further the P\'eclet number (Fig.~\ref{fig:3}(c)), the open chains signal disappears, showing an increase in the number of loops and the spiral chains. Consequently, the distribution function $P(\psi)$ exhibits four peaks: two for $1<\vert\psi\vert<2$, corresponding to metastable loops, and two for $2<\vert\psi\vert<3$, indicating stable spirals. On the other hand, for higher P\'eclet numbers, all filaments in the system are spirals, resulting in a probability distribution with only two symmetric peaks corresponding to a stable spiral phase (Fig.~\ref{fig:3}(d)). 

Finally, the distribution corresponding to $Pe=2.5 \times 10^4$ resembles that of $Pe=10^3$ (Fig.~\ref{fig:2}(b)), but the chains present in this case are much more compact. Additionally, note that in the dilute case ($\rho \leq 0.3$), just before the spiral transition, the probability distribution of the turning number resembles that observed for $Pe=2.5 \times 10^4$ (Fig.~\ref{fig:2}(e)). The latter implies that the open chain and spiral states are equally probable. These findings are consistent with those observed in the single filament example \cite{khalilian2024structural}, with further details available in the Supplementary Material~\cite{SI}. 

\begin{figure}[t!]
    \centering
        \includegraphics[width=8.5cm]{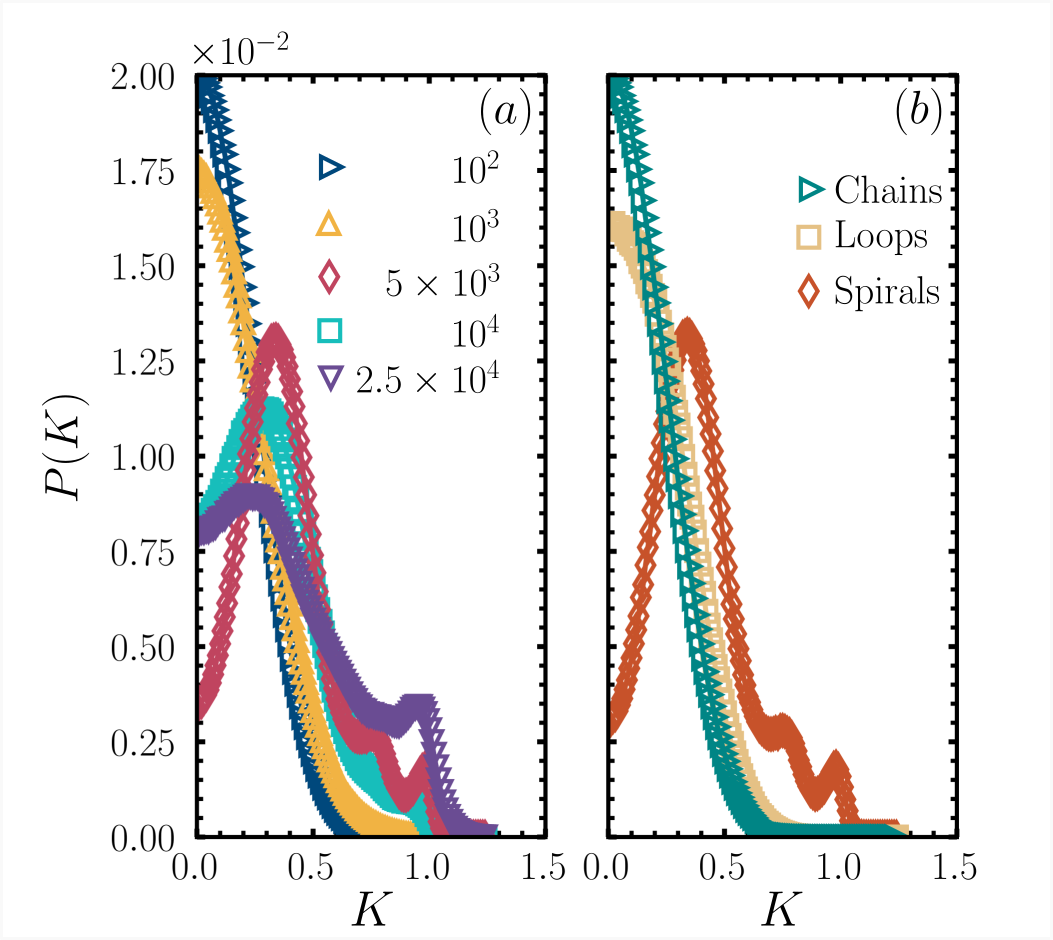}
     \caption{Probability distribution function of the curvature $P(K)$ at density $\rho=0.4$. (a) Each color and symbol corresponds to a different P\'eclet number $Pe$: blue right triangles for $Pe=10^2$, yellow triangles for $Pe=10^3$, cyan squares for $Pe=5.0 \times 10^3$, red diamonds for $Pe=10^4$, and purple upside-down triangles for $Pe=2.5 \times 10^4$. (b) Right green triangles represent the probability of the curvature for chains, beige squares correspond to loops, and dark orange diamonds correspond to spirals. The peak of $P(K)$ is consistent for open chains and loops but shifted to the right for spirals. At low P\'eclet numbers ($Pe \leq 10^3$), the probability distribution displays a single peak at $K \approx 0$,  reflecting the prevalence of stable open chains. For $10^3<Pe<10^4$, the distribution represents an average between loops and spirals, as both are present. Moreover, for $Pe=10^4$, the distribution aligns with that of spirals since all the filaments are spirals. Lastly, for $Pe>10^4$, the distribution $P(K)$ is an average between the distribution of chains and spirals, as both are present in the system. However, since the chains at this $Pe$ are more compact than those at $Pe=10^3$, the distributions at these two $Pe$ differ.}
     \label{fig:4}
\end{figure}

Because of the parallels between very high and smaller P\'eclet numbers, we can conclude that this distribution function alone does not provide significant information about the polymers with $\psi=0$. To discriminate between these types of chains, an additional measurement is required, which leads us to compute the curvature,
\begin{equation}
    K=\frac{x' y''-y'x''}{(x'^2+y'^2)^{\frac{3}{2}}},
\end{equation}
where $x$ and $y$ represent the coordinates of the beads, and the primes denote derivatives. Figure ~\ref{fig:4} shows that the curvature of the polymers changes substantially for different polymer shapes, making this measurement a valid parameter to distinguish between the different phases observed in the system.

Figure~\ref{fig:4}(a) and (b) show that the probability distribution $P(K)$ for small P\'eclet number is similar to the distribution observed for open chains. This correlation stems from the fact that, as discussed in the preceding section, at this P\'eclet number, all filaments in the system take the form of open chains. Furthermore, when $Pe$ increases, the system becomes more diverse, including loops and spirals. Therefore, the distribution $P(K)$ is an average of $P(K)$ corresponding to loops and spirals, as seen in Fig.~\ref{fig:4}(b).
In the spiral phase ($Pe=10^4$) the probability distribution peaks around $K \approx 0.5$, comparable to what is found for logarithmic spirals with a number of points equal to the number of beads $N_b$ and the same turning number $\vert\psi\vert$. Subsequently, for higher $Pe$ ($Pe=2.5 \times 10^4$), the probability distribution $P(K)$ diverges significantly from that observed just before the spiral transition ($Pe=10^3$). These results highlight the importance of filament curvature as a crucial metric for distinguishing between the various phases present in the system.

\subsubsection*{Spiral Characterization}
\begin{figure}[t!]
    \centering
    \includegraphics[width=8.5cm]{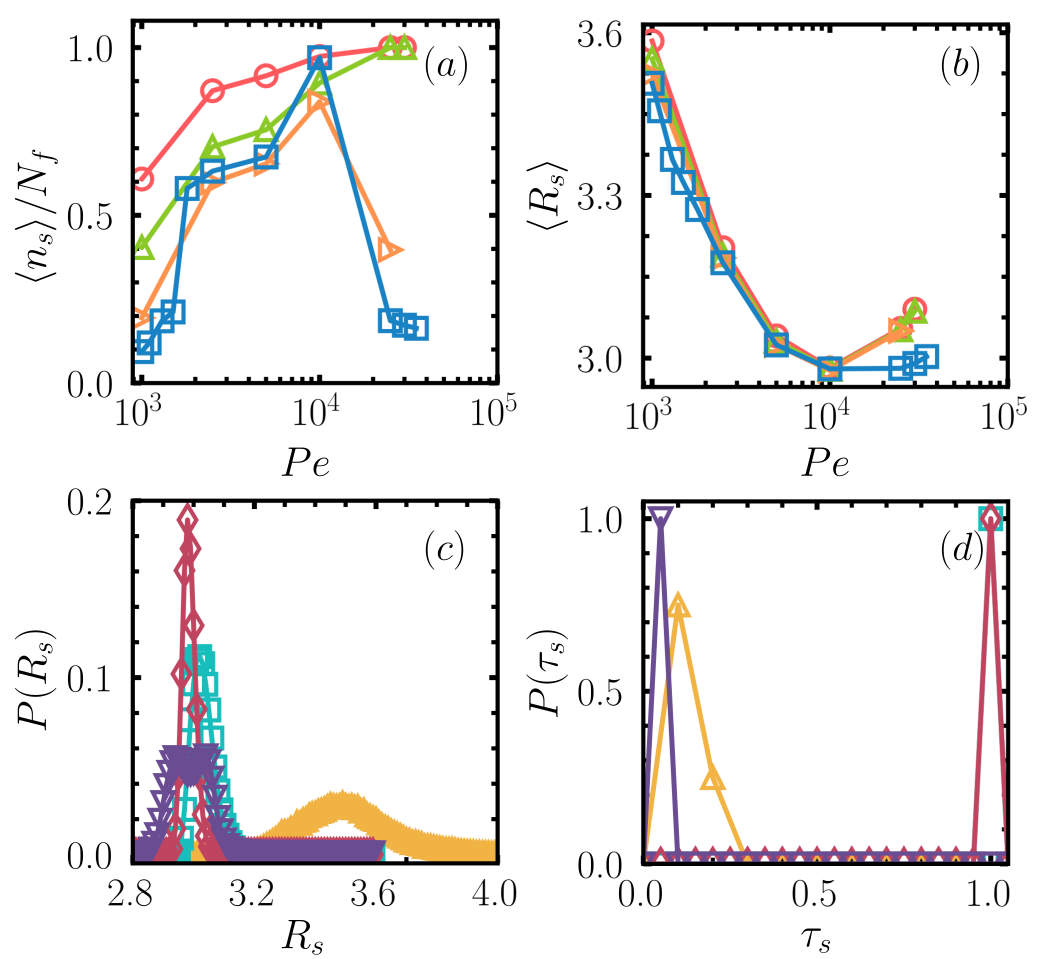}
     \caption{Spiral Characterization. In panels (a) and (b), different colors and symbols represent distinct densities: red dots for $\rho=0.2$, green triangles for $\rho=0.30$, orange right triangles for $\rho=0.35$, and blue squares for $\rho=0.40$. (a) Fraction of spirals present in the system $\langle n_s \rangle /N_f$ as a function of the P\'eclet number $Pe$. At low densities ($\rho \leq 0.30$), the number of spirals increases monotonically with $Pe$ until $Pe \approx 10^4$, where it plateaus to $\langle n_s \rangle /N_f \approx 1$ due to all filaments being spirals. At intermediate densities ($0.3 < \rho \leq 0.4$), a similar trend is observed, with the fraction of spirals increasing with $Pe$ until $Pe=10^4$, where it reaches a maximum at $\langle n_s \rangle /N_f \approx 1$ before decreasing due to a reentrant phase transition for $Pe>10^4$. (b) Gyration radius $\langle R_s \rangle$ of the spirals in the system as a function of $Pe$. For all densities, the gyration radius decreases with increasing $Pe$, indicating that spirals become more compact. Additionally, for $Pe>10^4$, the gyration radius slightly increases again due to the interaction between filaments, resulting in more logarithmic spirals. In panels (c) and (d), different colors and symbols represent different P\'eclet numbers $Pe$: yellow triangles for $Pe=10^3$, cyan squares for $Pe=5.0 \times 10^3$, red diamonds for $Pe=10^4$, and purple upside-down triangles for $Pe=2.5 \times 10^4$. (c) Probability distribution of the gyration radius of the spirals $P(R_s)$ at density $\rho=0.4$. This distribution illustrates that as $Pe$ increases, the spirals become increasingly compact. (d) Probability distribution of the normalized mean lifetime of the spirals $P(\tau_s)$ at density $\rho=0.4$. For $Pe=10^3$, spirals are unstable; for $10^3<Pe \leq 10^4$, spirals are stable, remaining in the spiral state without opening up throughout the entire measurement; and for $Pe>10^4$, the high P\'eclet number and the excluded volume effects cause spirals to open up and become unstable. Note that the legends for panels (a) and (b) correspond to Fig. \ref{fig:2}; and for panels (c) and (d) correspond to Fig. \ref{fig:4}}.
     \label{fig:5}
\end{figure}

Next, we analyze how the structure and stability of the spirals change across different P\'eclet numbers by thresholding the turning number to $\vert\psi\vert \geq 2.5$. At low density, the fraction of spirals increases with higher P\'eclet numbers until it reaches a plateau around $\langle n_s\rangle/N_f\approx 1$, indicating a scenario where all filaments in the system are spirals (Fig.~\ref{fig:5}(a)). Additionally, at intermediate densities, the fraction of spirals increases with $Pe$ up to $Pe\approx 10^4$. For larger P\'eclet numbers, however, the fraction of spirals decreases as the filaments unfold.
This behavior is in agreement with that observed for the end-to-end distance or the turning number (see Fig.~\ref{fig:2}).

To assess whether the spirals become more compact with increasing $Pe$, we calculate the gyration radius $R_s$ of the spirals,
\begin{equation}
    R_s=\sum_i (x_i-x_{cm})^2+\sum_i(y_i-y_{cm})^2
\end{equation}
where $\mathbf{r}_i=(x_i,y_i)$ is the instantaneous position of the i-\textit{th} monomer, and $\mathbf{r}_{cm}=(x_{cm},y_{cm})$ denotes the center of mass of the spiral. At low density, $R_s$ consistently decreases increasing $Pe$ until reaching $Pe=10^4$. This trend signifies progressive compactness in the spirals as the P\'eclet number increases. However, beyond $Pe=10^4$, a slight increase in the gyration radius is observed, due to the logarithmic shape of the spirals. On the contrary, as the density increases, the gyration radius diminishes as $Pe$ increases, and spirals become progressively compact (Fig.~\ref{fig:5} (c)).

Finally, to assess the stability of spirals at intermediate densities, we compute the duration for which a filament remains in the spiral configuration across multiple timeframes (Figure~\ref{fig:5} (d)). Around the P\'eclet number where the spiral phase is shown (Fig.\ref{fig:3}(b)), the spirals repeatedly form and dissolve numerous times within the measurement timeframe. As the P\'eclet number increases  (Fig.\ref{fig:2} (c),(d)), the spiral lifetime also increases. Filaments in the spiral configuration at the start of the measurement persist in this state throughout the measurement period $t_{s}$. Lastly, with further increasing the P\'eclet number (Fig.\ref{fig:2}(e)), at the reentrant phase, the spirals become highly unstable.
\begin{figure*}
    \centering
    \includegraphics[width=17cm]{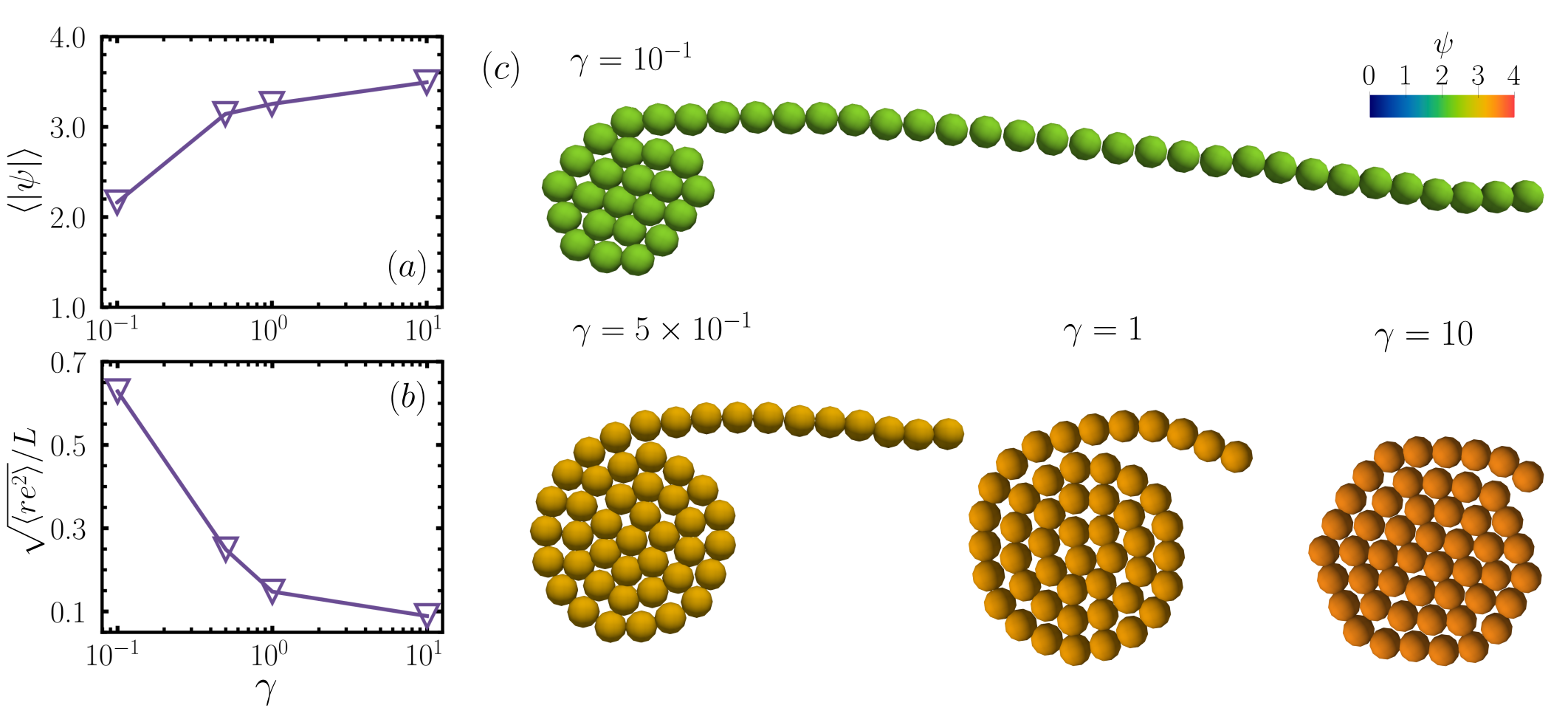}
    \caption{Effects of inertia on the structural properties of a single-polymer. The end-to-end distance (a) and turning number (b), are computed at $Pe=2.5 \times 10^4$, and $\gamma=0.1,0.5,1,10$. Both properties show that as $\gamma$ decreases the spirals become less compact. (c) Typical spiral shapes for different damping coefficients $\gamma$, with color coding based on the turning number.}
    \label{fig:6}
\end{figure*}

In summary, we have observed that the probability distributions of the turning number and curvature are effective metrics for distinguishing between open chains, compact chains, loops, and spirals. 
Our findings show that spirals become more compact with increasing P\'eclet number, and they exhibit stability in the pure spiral phase, while they are highly unstable at the onset of the spiral phase and within the reentrant phase.

\subsection{Inertial Effects}
We seek to gain a better understanding of the origins of the reentrant phase by investigating the role of inertial effects. Recently, three-dimensional simulations by ~\citeauthor{Fazelzadeh2023}~\cite{Fazelzadeh2023} revealed in the underdamped case, polymers tend to unwind.
Consequently, since the reentrant phase occurs within the regime where inertial effects are predominant ($Pe=2.5 \times 10^4$, i.e., $f_p=1.4$), we attribute this phase transition to inertial effects. To fully characterize the role of inertia, we start by analyzing a single polymer at $Pe=2.5 \times 10^4$ across various reduced time $\tau_m={m}/{\gamma}$. We maintain a fixed mass of one while varying $\gamma$.

Figure~\ref{fig:6}(a), and (b) shows the end-to-end distribution, and the turning number for different $\gamma$, respectively. With increasing $\gamma$ (or decreasing $\tau_m$), the spirals become progressively more compact. Additionally, at high friction coefficients ($\gamma=10$), approaching the overdamped limit, the probability of finding open chains in the system goes to zero. Here, the inertial timescale is significantly smaller compared to the timescale of the active force. Consequently, it is expected that the chains cannot unfold since, in this regime, inertial effects are not predominant~\cite{Fazelzadeh2023}. Conversely, as $\gamma$ decreases (or $\tau_m$ increases), exceeding the timescale of the active force, the spirals become less compact, and the probability of finding open chains increases(see Supplementary Material~\citep{SI}). 
Moreover, when inertial effects are negligible, the spirals observed in the system are very compact and can be accurately fitted with Archimedean spirals(Fig.~\ref{fig:6}(c)). As the friction coefficient $\gamma$ decreases, however, the spiral's end becomes increasingly detached from the center of mass, resulting in an elongated arm. This behavior change is attributed to the predominance of inertial effects and the bending potential. Note, that in a recent preprint, it has been shown that inertial effects in a two-dimensional single polymer system lead to a transition from spinning Archimedean spirals to semi-flexible open chains with increased activity, without observing changes in the types of spiral structures~\cite{karan2024inertia}. However, this difference might be due to the smaller range of $\tau_m$ used in their simulations.

\begin{figure}[t!]
    \centering
    \includegraphics[width=8.5cm]{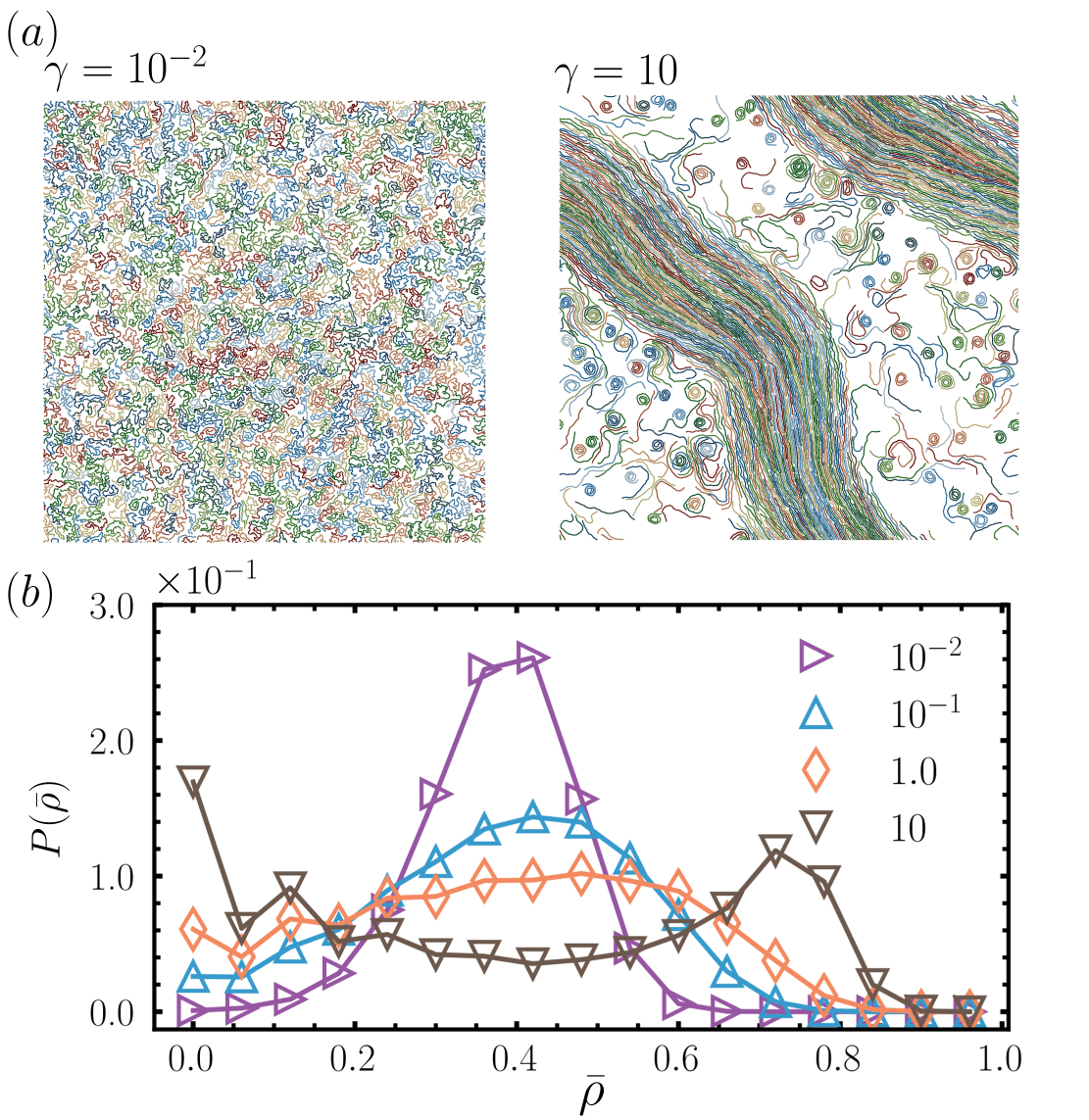}
    \caption{Interplay of inertia with density. (a) Snapshots for $\gamma=10^{-2}, 10$. (b) The probability distribution of coarse-grained local density $P(\bar{\rho})$ at $Pe=10^{3}$ for various $\gamma$ values. Phase separation systems have a two-peak distribution curve for large $\gamma$ values. The coarse-grained local density $\bar{\rho}$ shows a single peak for $\gamma<1.0$, which corresponds to the actual system density as expected. }
    \label{fig:7}
\end{figure}

Finally, we investigated the interplay of inertia with density, specifically focusing on the case where the reentrant phase is observed ($\rho=0.4$). At low P\'eclet numbers and intermediate $\gamma$, a dense system experiences phase separation, referred to as a flowing melt~\cite{Prathyusha2018}. As $Pe$ increases, filaments become uniformly distributed. In contrast, as illustrated in~\ref{fig:7}(a), with increasing $\gamma$ (approaching the overdamped limit), the system undergoes phase separation. The disappearance of the phase separation at high P\'eclet numbers and low damping coefficients is attributed to the fluctuations induced by inertia, causing polymers to collide and rebound~\cite{suma2014motility, mandal2019motility, Loweninertia2020}.
To investigate phase separation, we compute the coarse-grained local density $\bar{\phi}(\mathbf{r})$. Following \citeauthor{testard}~\cite{testard}, we discretize the system by dividing it into squares of size $d$ so that continuous space is now replaced by a discrete lattice containing $(L/d)^2$ sites. Then, we construct the coarse-grained local density, $\bar{\phi}(\mathbf{r})$, defined for discrete positions $\mathbf{r}$ located at the center of each grid square given by
\begin{equation}
    \bar{\rho}(\mathbf{r}) = \frac{\sigma^2}{4d^2} \sum_{i=1}^N \Theta(d - \vert \mathbf{r}-\mathbf{r}_i\vert),
\end{equation}
 where $\Theta(y)$ is the Heaviside function, and $d=4\sigma$. 
 
When $\gamma$ is extremely small and inertial effects are dominant, the probability distribution of the coarse-grained local density $P(\bar{\rho})$ exhibits a single peak at values of the local density $\bar{\rho}$ corresponding to the system's density $\rho$ (Figure~\ref{fig:7}(b)). This suggests a uniform distribution of filaments across the system. As the damping coefficient increases, the peak of $\bar{\rho}$ broadens until it reaches a critical value ($\gamma=10$), where $P(\bar{\rho})$ shows numerous peaks, confirming the presence of phase separation in the system. This behavior is consistent with what is observed in MIPS Ref.~\cite{mandal2019motility}. In our case, large values of  $\gamma$, shift in the spiral transition to higher $Pe$. However, as $\gamma$ decreases, not only is the phase-separation phase absent, but the spiral phase also vanishes. In this case, the inertial effects are so pronounced that spiral formation is inhibited. 
 
Overall, our findings suggest that inertia influences the shape of spirals in a single polymer. Moreover, in a dense system, inertia plays a pivotal role in shifting the spiral transition to higher $Pe$ when inertial effects are negligible and vanishing the spiral transition when inertial effects are substantial. These findings suggest that the presence of the reentrant phase is attributable to both the interactions between polymers and inertial effects.

\section{Conclusions}
In summary, our study highlights the importance of density and inertia on the behavior of active polymers. Remarkably, at intermediate densities and damping coefficient, the system exhibits a reentrant phase. At low P\'eclet numbers, the system manifests as a polymer melt phase,  transitioning to a pure spiral phase at high P\'eclet numbers. As the P\'eclet number increases even more, the spirals become unstable, leading to their unwinding. This reentrant phase does not appear in the dilute \cite{IseleHolder2015} or the high-density limit \cite{Prathyusha2018}. In the dilute case, the fact that spirals do not unwind is because excluded volume effects are not relevant. As density increases, the system transitions from a polymer melt phase to a state characterized by open chains and unstable spirals. The absence of a reentrant phase can be attributed to the lack of a phase where all filaments are in a spiral configuration. The absence of the pure spiral phase is due to excluded volume effects, i.e., as density rises, there is insufficient space for all filaments to be in a spiral configuration.

To gain a deeper understanding of the cause of this reentrant phase, we have also investigated the impact of inertia on the system's behavior.
In the case of a single-polymer, our findings revealed that underdamped conditions result in spirals that progressively unwind as the damping coefficient decreases \cite{Fazelzadeh2023}. This leads to spirals with an arm increasingly distant from their center of mass. Conversely, increasing the damping coefficient causes spirals to become more compact, resembling Archimedean spirals. Furthermore, we explored how these inertia effects influence the behavior of polymer suspensions showing that inertia contributes to the occurrence of the reentrant phase. As the P\'eclet number rises, the spirals become less compact, and due to the excluded volume they unwind. When inertial effects are minimal (e.g., high damping coefficients), the system undergoes phase separation ~\cite{mandal2019motility} even at high P\'eclet numbers, resulting in a shift in the spiral transition. Therefore, the observed reentrant phase at intermediate values of the damping coefficient arises from a non-trivial interplay between density and inertial effects.

Our study provides new insights into the rich non-equilibrium dynamics of active polymers, underlining the importance of investigating the interplay between density and inertia. Considering the increasing relevance of dense active matter in biological contexts, where inertia also plays a role, our findings hold potential applications in such domains. Spiral formation has not yet been observed in experimental studies~\cite{Sumino2012, IseleHolder2015}. Our investigation revealed that under underdamped conditions in a system of polymer suspensions, the system does not exhibit a spiral phase. This suggests that the absence of spirals in experiments may be attributed to both inertial effects and interactions among different polymers.

\section*{Acknowledgements}
D.M.F. and GJ thanks the Comunidad de Madrid and the Complutense University of Madrid (Spain) through the Atraccion de Talento program 2022-T1/TIC-24007. D.M.F. would like to thank Sara Jabbari-Farouji for the insightful discussions and valuable feedback that greatly enriched this research. 
\bibliography{./biblo} 
\bibliographystyle{apsrev4-1} 
\end{document}